\documentclass[preprint,12pt]{aastex}

\shorttitle{Orbit of Sagittarius}
\shortauthors{Newberg, Yanny et al.}

\begin{document}

\title{Sagittarius Tidal Debris 90 kpc from the Galactic Center}

\author{
Heidi Jo Newberg\altaffilmark{\ref{eq}, \ref{RPI}},
Brian Yanny\altaffilmark{\ref{eq}, \ref{FNAL}},
Eva K. Grebel\altaffilmark{\ref{MPH}},
Greg Hennessy\altaffilmark{\ref{usnoDC}},
\v{Z}eljko Ivezi\'{c}\altaffilmark{\ref{PU}},
David Martinez-Delgado\altaffilmark{\ref{MPH}},
Michael Odenkirchen\altaffilmark{\ref{MPH}},
Hans-Walter Rix\altaffilmark{\ref{MPH}},
Jon Brinkmann\altaffilmark{\ref{APO}},
Don Q. Lamb\altaffilmark{\ref{UC}},
Donald P. Schneider\altaffilmark{\ref{PSU}}
Donald G. York\altaffilmark{\ref{UC}}
}

\altaffiltext{1}{Equal first authors\label{eq}}

\altaffiltext{2}{Dept. of Physics, Applied Physics and Astronomy, Rensselaer
Polytechnic Institute Troy, NY 12180; heidi@rpi.edu\label{RPI}}

\altaffiltext{3}{Fermi National Accelerator Laboratory, P.O. Box 500, Batavia,
IL 60510; yanny@fnal.gov; skent@fnal.gov\label{FNAL}}

\altaffiltext{4}{Max Planck Institute for Astronomy, K\"onigstuhl 17, D-69117 Heidelberg, Germany \label{MPH}}

\altaffiltext{5}{US Naval Observatory, 3450 Massachusetts Avenue NW, Washington, DC 20392-5420 \label{usnoDC}}

\altaffiltext{6}{Princeton University, Princeton NJ 08544\label{PU}}

\altaffiltext{7}{Apache Point Observatory, P. O. Box 59, Sunspot, NM 88349-0059\label{APO}}

\altaffiltext{8}{Dept. of Astronomy and Astrophysics, University of Chicago, 5640 S. Ellis Ave., Chicago, IL 60637\label{UC}}

\altaffiltext{9}{Department of Astronomy and Astrophysics, The Pennsylvania State University, University Park, PA 16802\label{PSU}}

\begin{abstract}

A new overdensity of A-colored stars in distant parts of the Milky Way's stellar halo, 
at a dereddened SDSS magnitude of $g_0 = 20.3$, is presented.  Identification of
associated variable RR Lyrae candidates supports the claim that these are blue
horizontal branch stars.  The inferred distance of
these stars from the Galactic center is 90 kpc, assuming the absolute magnitude
of these stars is $M_{g_0} = 0.7$ and that the Sun is 8.5 kpc from the Galactic
center.  
The new tidal debris is within 10
kpc of same plane as other confirmed tidal debris from the disruption of the 
Sagittarius dwarf galaxy, and could be associated with the trailing tidal arm.
Distances to the Sagittarius stream estimated from M stars are about $13\%$
smaller than our inferred distances.  
The tidal debris has a width of at least $10^\circ$, and
is traced for more than $20^\circ$ across the sky.  The globular cluster NGC 2419
is located within the detected tidal debris, and may also
have once been associated with the Sagittarius dwarf galaxy.

\end{abstract}


\section{Introduction\label{intro}}

The discovery of the Sagittarius dwarf galaxy \citep{igi94}, which is in the process
of being tidally disrupted by the gravitational field of the Milky 
Way galaxy, has sparked a decade of research on the nature, extent, and
formation history of its tidal tails.  Underlying the quest for the details of this
merger event in our own galaxy is the hope that it will lead to a better understanding
of halo formation processes in general and the formation of the Milky Way galaxy in
particular.  Additionally, the tidal streams can be used as a probe of the Galactic
potential, and the clumpiness of the halo dark matter component.

In this letter, we present evidence that a piece of the Sagittarius dwarf tidal 
stream extends to a distance of up to 90 kpc from the Galactic Center.  This doubles the
known extent of the trailing tidal tail.  These stars will help constrain models of the
Galactic potential and the disruption of the Sgr dwarf.

\section {Observations and Data}

The SDSS scans the sky in 2.5 degree wide stripes that follow great circles on the sky.  
Details of the survey geometry may be found in \cite{setal01} and \cite{aetal03}.
Other SDSS technical details can be found in: \citet{figdss96},
\citet{getal98}, \citet{hfsg01}, \citet{pmhhkli03}, \citet{setal02}, and \citet{yetal00}.

SDSS imaging data available 2003 March
was searched for clumps of faint blue stars, following the techniques
described in \cite{ynetal00}, hereafter Paper I, for selecting stars of type A.
These stars are bluer than the turnoff of the Galactic thick disk,
thereby minimizing contamination from this component.
A-type stars (including many Blue Horizontal Branch - BHB - objects), are selected
with $(u-g)_0 > 0.4$ (to avoid contamination from extragalactic quasars) and $-0.3 < (g-r)_0 < 0$.
This `standard candle' population was used to explore the stellar halo 
structure out to distances of $\sim 100$ kpc from the Sun.  

Table 1 lists the SDSS stripes examined for 
which detections near the plane of the Sgr dwarf orbit \citep{ilitq01} were noted.  
Listed are the stripe number, stripe inclination, Galactic latitude and longitude of the center 
of intersection of the Sagittarius stream with the observed SDSS stripe,
mean $g$ magnitude of the BHB stars in the stream, a derived distance to
the stream stars, and rough number counts, $N$, of type A stars in each stripe 
which crosses the stream, along with the angle between the A star distribution and
the end of the stripe.  Note that the stripes overlap and thus the numbers may 
not be summed without accounting for the overlap.  About 20\% of the BHBs are listed
twice in the Table, due to the overlap of adjacent stripes.  
The number counts of A stars were determined by counting the number of A stars within
0.2 magnitudes of the tabulated $g$ magnitude, and within $20^\circ$ of the tabulated
position.  A reasonable background A star count was estimated from a histogram of
apparent magnitudes in each direction and has been subtracted.
We note qualitatively that the spatial density of BHB stars remains approximately as 
high at 90 kpc from the Galactic Center (stripes 27 through 37) as it does 
for trailing tail stripes 82 and 86, only 30 kpc from the G.C. 

\section{Detection of a new density enhancement in the Galactic halo}

Figure \ref{arcs} shows a polar wedge (magnitude, angle) diagram for type A/BHB
objects in the SDSS stripe 29.  The most striking new feature
is the overdensity of A-type stars at $g_0 = 20.3$
(where the subscript indicates the magnitude has been dereddened by the 
standard \cite{sfd98} maps) and $(l, b) = (195^\circ,29^\circ)$.
There is another apparent overdensity of A-type stars with $g_0 \sim 17.5$ (apparent also in Figure \ref{trail90}).  The
source of these stars is unclear.  They could be
blue stragglers in the new structure discovered in the plane of the
Milky Way \citep{ynetal03}.  Alternatively, they could be associated with
a Galactic disk structure or a closer piece of debris from the Sgr dwarf
galaxy.

Figure \ref{trail90} shows the $g_0$ magnitude of all A-type stars in stripes 27-37 with
$100^\circ < \alpha < 170^\circ$ and within 15 kpc of the Sgr dwarf tidal plane, as defined in \citet{mswo03} 
(see \S 4), as a function of $\Lambda_\odot$.  $\Lambda_\odot$ was defined by \citet{mswo03}
to be the angular distance from the Sgr dwarf, in the plane of the Sgr dwarf orbit.
This figure can be compared with Figure 8 of \citet{mswo03}.  From Figure \ref{trail90},
this tidal fragment (at $g_0 \sim 20.3$) extends at least $20^\circ$ on the sky ($-173^\circ < \Lambda_\odot < -148^\circ$).
If one adjusts for the fact that the stripe in Figure \ref{arcs} is not perpendicular to the Sgr
orbit (which makes the stream appear wider), we find that the stream is at least 10 degrees
across.  When this overdensity was discovered,
it was not in a targeted search for stars in the Sgr stream, but it was the most significant
new substructure discovered in a routine search through the 
entire SDSS data set collected to date.

Since this identified overdensity is 
narrowly concentrated in apparent magnitude, we tentatively identify this as a population of
blue horizontal branch (BHB) stars associated with a tidal stream in the outer halo.  Using an
absolute magnitude of $M_g \sim 0.7$ (Paper I), the tidal stream is 83 kpc from the Sun and
at least 15 kpc in width.  With a position $(l, b) = (190^\circ, 30^\circ)$, the overdensity 
is 90 kpc from the Galactic center.

Note the `finger of God' at
$(\Lambda_\odot, g_0) = (-170^\circ, 20.3)$.  These are the horizontal branch stars from the globular
cluster NGC 2419 ($[l,b] = [180^\circ, 25^\circ]$).  
The magnitude of the horizontal branch is less than a tenth of a magnitude different from those of
our newly identified distant tidal stream.  NGC 2419 is 13 kpc above the Sgr dwarf orbital
plane.
This cluster could have once belonged to the Sgr 
dwarf galaxy or some parent progenitor to today's Sgr dwarf.
The fact that this cluster is relatively close the the plane of the Sgr dwarf tidal debris
has also been noticed by \citet{z98}.

The globular cluster NGC 2419 has a heliocentric radial velocity of -20 km/s \citep{poa86}.  This
translates to a radial velocity of -14 km/s in the rest frame of the Galactic center, but as seen from
the Sun.  Since the Sun and the Galactic center are only about $2^\circ$ apart as seen from
NGC 2419, the radial velocity as seen from the Sun should be close to the radial velocity as measured
from the Galactic center.  If the globular cluster is in a portion of the stream that is near
apogalacticon, as suggested by our results, the galactocentric radial velocity should 
be near zero.  With a velocity dispersion of 20 km/s
\citep{ynetal03}, the observed radial velocity is consistent with its association with the 
Sgr dwarf tidal stream near its apogalacticon.

We have confirmed the identification of the A-colored stars as horizontal branch stars by identifying
an excess of RR Lyrae candidates with the correct apparent magnitudes in this position.  
We selected stars in the overlaps of stripes 27 through 37 which were observed at multiple epochs.
Ordinarily this SDSS stripe-to-stripe overlap is about 20\% of the 
area of the sky.  However, since these observations are near the beginning and ends of
SDSS survey stripes (near the survey poles) the stripes converge, 
significantly increasing the overlap.  The matching routine first selected objects
with $16 < g_0 < 22$ in the color range $-0.2 < (g-r)_0 < 0.3$, and chose stellar objects at the 
same spatial position (within $1''$) from the list which were observed more than once.
We compared the matches in the area $110^\circ < \alpha < 130^\circ$,  $20^\circ < \delta < 50^\circ$ 
with a wider `control' area with a similar number of matches over the whole of the stripes covering 
$165^\circ < \alpha < 260^\circ, 20^\circ < \delta < 65^\circ$.  

There are about 50 excess variable objects with magnitude differences of at least 0.4 magnitudes 
between the two observations in the magnitude range $20 < g < 21$ (57 in the first region vs. 5 in the control region).
This sample of RR Lyrae candidates provides strong confirmation
that the blue A-type stars seen in Figure \ref{arcs} and \ref{trail90} near the
plane of the Sgr dwarf orbit, some 81-85 kpc from the Sun, are indeed
BHB stars and not intrinsically fainter (and thus more nearby) blue straggler
objects of the same color (Paper I).

\section{The trailing tail of the Sagittarius tidal stream}

We compare our detection with the positions of the Sgr tidal tails of 2MASS M-stars 
as presented by \citet{mswo03}.  In order to put our data on their Figure 11, showing the
tidal tails, we adopt their distance from the Sun to the center of the Galaxy of 8.5 kpc. 
Our Galacitic X-axis is along the line from the Sun to the center of the Galaxy, but 
incontrast to \citet{mswo03} it is measured from the Galactic center 
towards $l=0^\circ$, so that the Sun is at $X = -8.5$ kpc.  
$Y$ ($l = 90^\circ$) and
$Z$ ($b = 90^\circ$) are unchanged.  Using our definitions, the Sgr plane is:

$-0.064 X + 0.970 Y + 0.233 Z + 0.232 = 0.$

The same paper defines another coordinate system, $(X_{Sgr, GC}, Y_{Sgr, GC}, Z_{Sgr, GC})$, where
$Z_{Sgr, GC}$ is perpendicular to the plane of the orbit, and $Y_{Sgr, GC} = 0$ in the Galactic plane.
Figure \ref{aboveplane} shows the 2MASS M stars from Figure 11 of \citet{mswo03}, along with 
the positions (large open circles) of the
stream we detect in SDSS A-stars in stripes 29-37.  Also shown are other detections of the tidal tails
of Sgr in SDSS A-stars, by methods described in \citet{netal02}. 
Open circles denote the trailing
tail, and filled circles mark the leading tail.  The error bars are 
three sigma error bars, but do not include systematic error which would arise if the absolute
magnitudes of our selected A stars were not $M_{g_0} = 0.7$.  We measure the central 
apparent magnitude of each overdensity to within 0.05 magnitudes in $g_0$, and the 
position along the stripe of data to within one
degree.  We added in quadrature the errors in distance from the center of the Galaxy that would arise 
from a difference of 0.15 magnitudes in apparent magnitude and from a difference of $3^\circ$ in 
stream center.  The data points for the distant portion of the stream were determined from
the entire dataset of A-type stars converted to $\Lambda_0, g_0$
coordinates; their positions are not tied to individual SDSS stripes, but spaced according to the
density of stars in Figure \ref{trail90}, excluding NGC 2419.  

There appears to be a systematic scale difference between the distance to 
the Sgr stream from 2MASS M star statistics and from A star statistics.  The A star distance
scale used in Paper I was based on the somewhat arbitrary assumption that the blue
horizontal branch (BHB) stars in the sample had absolute magnitudes of $M_{g_0} = 0.7$.  Note, however, that the distances
given in Paper I to pieces of the Sgr tidal stream from 
BHB stars are not far from the distances quoted in \citet{ietal00} and \citet{vetal01} to similar sections of the
stream using RR Lyrae stars.  
The nominal distance to NGC 2419 is 81 kpc \citep{hal97} from the Sun, in excellent agreement
with our measured distance to the tidal stream.
%One delta Scuti star at $(l,b) = (174^\circ, -51^\circ)$ was selected from the SDSS database and confirmed
%with the USNO 1m in Flagstaff, AZ.  From the observed lightcurve, the inferred distance to this star, which is in 
%the same direction as the Sgr stream BHBs observed in SDSS stripe 82, is $37\pm2$ kpc, which if 
%all the assumptions are true would push towards a larger distance scale 
%to the stream (Table 1 lists the BHB stars in this same direction to have an inferred distance of 29 kpc).
The M-star distance scale is calibrated assuming a distance of 24 kpc from the Sun
to the Sgr dwarf, 12.5\% smaller than the distance to M54 measured by \citet{ls00} using
RR Lyrae stars.  If the larger distance scales for A-stars and RR Lyraes are correct, then
either the M-star population changes as a function of position along the stream (as suggested by 
Martinez-Delgado 2003), or the Sgr dwarf is somewhat more distant than typically assumed.  
Alternatively, the absolute magnitudes
of the BHB and RR Lyrae stars could be $\sim 0.26$ magnitudes fainter than we have previously assumed.

The newly discovered overdensity of A-type stars may coincide with an overdensity of M giants
on the left edge of Figure 9 of \citet{mswo03}.  These stars appear to extend the trailing tail of the 
Sgr dwarf back up through the Galactic plane into the northern Galactic hemisphere.
Our result suggests that these M-stars may in fact be a part of a
longer trailing tail of the Sgr dwarf, which extends to 90 kpc from the Galactic center.

Figure \ref{planefit} shows the positions of the same pieces of tidal debris, but shown as a 
cross section through
the plane of the Sgr dwarf tidal stream.  The exact position of the center of the tidal stream
is uncertain for the distant portion of the tidal stream, since we do not unambiguously detect the entire
cross section in any stripe, and in many stripes we detect only a portion of the cross section.
The $3\sigma$ asymmetrical error bars reflect this fact.
If we adopted a larger absolute magnitude for the A-type stars, then position of the points on the X-axis
would shift somewhat but the positions and errors on the Z-axis would hardly change.  Notice that the
leading arm of Sgr shifts towards larger (positive) $Z_{Sag}$ as we go back 
towards the Sgr dwarf.  The trailing arm (closer than 50 kpc) is less well sampled, but possibly shifts towards 
smaller (negative) $Z_{Sag}$ as we go towards the Sgr dwarf.  This may indicate that 
the leading arm originates at positive $Z_{Sag}$ and the trailing
arm at negative $Z_{Sag}$.  The slight deviations from the plane might alternatively result from a
non-spherical Galactic potential.  These deviations cannot be reduced by redefining the Sgr orbital plane.

\section{Conclusions}

We have detected an overdensity of A-type stars with an estimated distance of 90 kpc from the
Galactic center.  The globular cluster NGC 2419 appears to be within this overdensity, which
is in nearly the same plane as the previously discovered
tidal tails of the Sgr dwarf galaxy.  The radial velocity of
NGC 2419 is consistent with its association with the tidal stream, near the apogalacticon of its
orbit.  The position of the
overdensity is consistent with identification as a distant piece of the trailing tail of the
Sgr dwarf galaxy, and coincident with the positions of Sgr dwarf M giants
discovered in 2MASS.  The inferred distances to the M giants are 13\% smaller than those
from BHB and RR Lyrae stars in the Sgr stream.  This may indication a systematic error
in one or both of the distance scales.  

\acknowledgments

HJN acknowledges funding from Research Corporation and the National Science Foundation 
(AST-0307571).   We are grateful to S. Majewski for permission to reproduce 
a figure of 2MASS M giant stars in the stream. We thank the referee, Helio J. Rocha-Pinto, for his kind suggestions.

Funding for the creation and distribution of the SDSS Archive has been
provided by the Alfred P. Sloan Foundation, the Participating
Institutions, the National Aeronautics and Space Administration, the
National Science Foundation, the U.S. Department of Energy, the
Japanese Monbukagakusho, and the Max Planck Society.  The SDSS Web
site is http://www.sdss.org/.

The SDSS is managed by the Astrophysical Research Consortium (ARC) for the 
Participating Institutions. The Participating Institutions are The University 
of Chicago, Fermilab, the Institute for Advanced Study, the Japan Participation 
Group, The Johns Hopkins University, Los Alamos National Laboratory, the 
Max-Planck-Institute for Astronomy (MPIA), the Max-Planck-Institute for 
Astrophysics (MPA), New Mexico State University, University of Pittsburgh, 
Princeton University, the United States Naval Observatory, and 
the University of Washington.

\clearpage

\figcaption {Polar wedge diagram of color-selected ($-0.3 < g_0 < 0$) A-type stars in stripe 29. 
The $g_0$ apparent magnitude is plotted 
radially, and the angle is angular position along the great
circle navigated by stripe 29.  The diagram has been oriented so
that the horizontal line at the base is the intersection of the plane
of stripe 29 and the Galactic plane.  
%Fainter than $g_0 = 21.5$,
%quasars are not effectively rejected by a color cut in $u_0-g_0$,
%and leak into the sample.
Notice the new overdensity of A-type stars at $g_0 = 20.3$ at the right side
of the figure.  The dashed line that defines the intersection of
the plane of orbit of the Sgr dwarf and the observation
plane. 
\label{arcs}} 
%goes near the center of the overdensity, suggesting
%that this new structure might be associated with the tidal debris
%from the Sgr dwarf.

\figcaption {Plot of A-colored stars in SDSS stripes 27 
through 37 ($100^\circ < \alpha < 170 ^\circ$).  The x-axis shows 
$\Lambda_\odot$, in degrees, which is the angle, in the
orbital plane of the Sgr dwarf galaxy, between each
measured A-type star and the dwarf.  It is the longitude, in a
coordinate system where plane of the Sgr dwarf tidal
debris has a latitude of zero degrees,
and is the same angle plotted in 
Figure 8 of \cite{mswo03}.  The y-axis gives the $g_0$ magnitude for the same
stars.  Notice the overdensity of stars at about $g_0 = 20.3$ extending from
$-148^\circ < \Lambda_\odot < -172^\circ$. This is a tidal tail of stars, close to the plane of the Sgr orbit, but at an implied distance of 
90 kpc from the Galactic center.
\label{trail90}}

\figcaption{The plane of the Sagittarius dwarf tidal debris.  The 2MASS M giants 
from Figure 11 of \cite{mswo03} are 
reproduced here, along with the position and direction of motion of the Sgr dwarf
galaxy (dark filled circle and line).  Superimposed on the M giants are detections of SDSS A-colored
stars in stripes 9-14, 27-37, 82 and 86 (points with error bars going 
counter-clockwise starting from the upper right).  
%The coordinate axes are $X_{SGR,GC}$ and $Y_{SGR,GC}$ 
%which are related to Galactic (X,Y,Z) as indicated.  
The Sun is at (X,Y,Z) = (-8.5,0,0) kpc. There is
a 13\% discrepancy in distance scale between 
the two data sets, reflecting uncertainties in the 
distance to Sgr and in the absolute magnitudes of A-colored
stars and M giants.  The four large open circles, showing the new tidal debris from stripes 27-37,
could be a continuation of the trailing tail of the Sgr
dwarf galaxy. \label{aboveplane}}

\figcaption{Cross section through the Sgr orbital plane.  We
show the positions of the stream centers of SDSS A-colored stars
perpendicular to the plane of Sgr taken from \cite{mswo03}.  While
the stars within 50 kpc of the Galactic center are very close to the plane, 
those at 90 kpc are as much as 10 kpc from the nominal orbital plane.
The rightmost open circle shows the approximate position of the Sgr dwarf galaxy, with
a nearly horizontal line showing the range of possible positions for
heliocentric distances between 21 and 29 kpc.
The leading tail (filled circles), trailing tail (open circles), and the newly
detected disrupted material (larger open circles) are also shown.
Note that the leading tail has a trend
towards positive $Z_{Sag, GC}$ as one goes towards the Sgr dwarf,
and the trailing tail tends towards more negative $Z_{Sag, GC}$ as one
goes towards the Sgr dwarf. \label{planefit}}

\clearpage
\begin{deluxetable}{rrrrrrrr}
\tabletypesize{\scriptsize}
\tablecolumns{8}
\footnotesize
\tablecaption{Table of SDSS Stripes with Sgr Stream BHBs}
\tablewidth{0pt}
\tablehead{
\colhead{Stripe} & \colhead{incl.\tablenotemark{1}} 
& \colhead{l} & \colhead{b} & \colhead{$g$} & \colhead{$d$\tablenotemark{2}} & \colhead{N} & \colhead{angle to edge}\\
\colhead{Number} & \colhead{$^\circ$} &  \colhead{$^\circ$} & \colhead{$^\circ$} & \colhead{mag} & \colhead{kpc} & \colhead{} & \colhead{$^\circ$}}

\startdata
9 &  -2.5 & 347.9& 51.4& 19.15 & 49 & 115 & 19\\
10 &  0.0 & 343.2 & 56.1 & 19.12 & 48 & 125 & 34\\
11 &  2.5 & 330.2 & 62.6 & 18.9 & 44 & 58 & 37\\
12 &  5.0 & 323.7 & 66.3 & 18.8 & 42 & 60 & 38\\
14 &  10.0&  284.0 & 72.0 & 18.6 & 38 & 25 & 30\\
27 &  42.5&  197.0 & 34.0 & 20.30 & 83 & 55 & 13 \\
28 &  45.0&  194.0 & 34.0 & 20.30 & 83 & 48 & 11  \\
29 &  47.5&  195.4 & 28.5 & 20.30 & 83 & 58 & 10 \\
30 &  50.0&  194.3 & 27.0 & 20.30 & 83 & 31 & 10\\
31 &  52.5&  191.7 & 28.1 & 20.30 & 83 & 39 & 11\\
32 &  55.0&  190.1 & 27.3 & 20.30 & 83 & 24 & 9\\
33 &  57.5&  188.7 & 26.5 & 20.30 & 83 & 45 & 9\\
34 &  60.0&  187.2 & 25.6 & 20.35 & 85 & 39 & 9\\
35 &  62.5&  184.4 & 26.1 & 20.35 & 85 & 31 & 7\\
36 &  65.0&  186.7 & 21.3 & 20.35 & 85 & 16 & 3\\
37 &  67.5&  182.5 & 23.2 & 20.35 & 85 & 113\tablenotemark{3} & 3\\
82 &   0.0&  163.2 & -56.1 & 18.0 & 29 & 35 &42 \\
86 &  10.0&  148.8 & -70.8 & 18.1 & 30 & 24 & 45\\
\enddata
\tablenotetext{1}{Inclination of stripe relative to celestial equator, node is at $\alpha = 95^\circ$.}
\tablenotetext{2}{Inferred distance of stars from the Sun, assuming $M_g(\rm BHB) = +0.7$.}
\tablenotetext{3}{Includes NGC 2419 BHB stars.}
\end{deluxetable}

\clearpage

\setcounter{page}{1}

\plotone{f1.ps}

\plotone{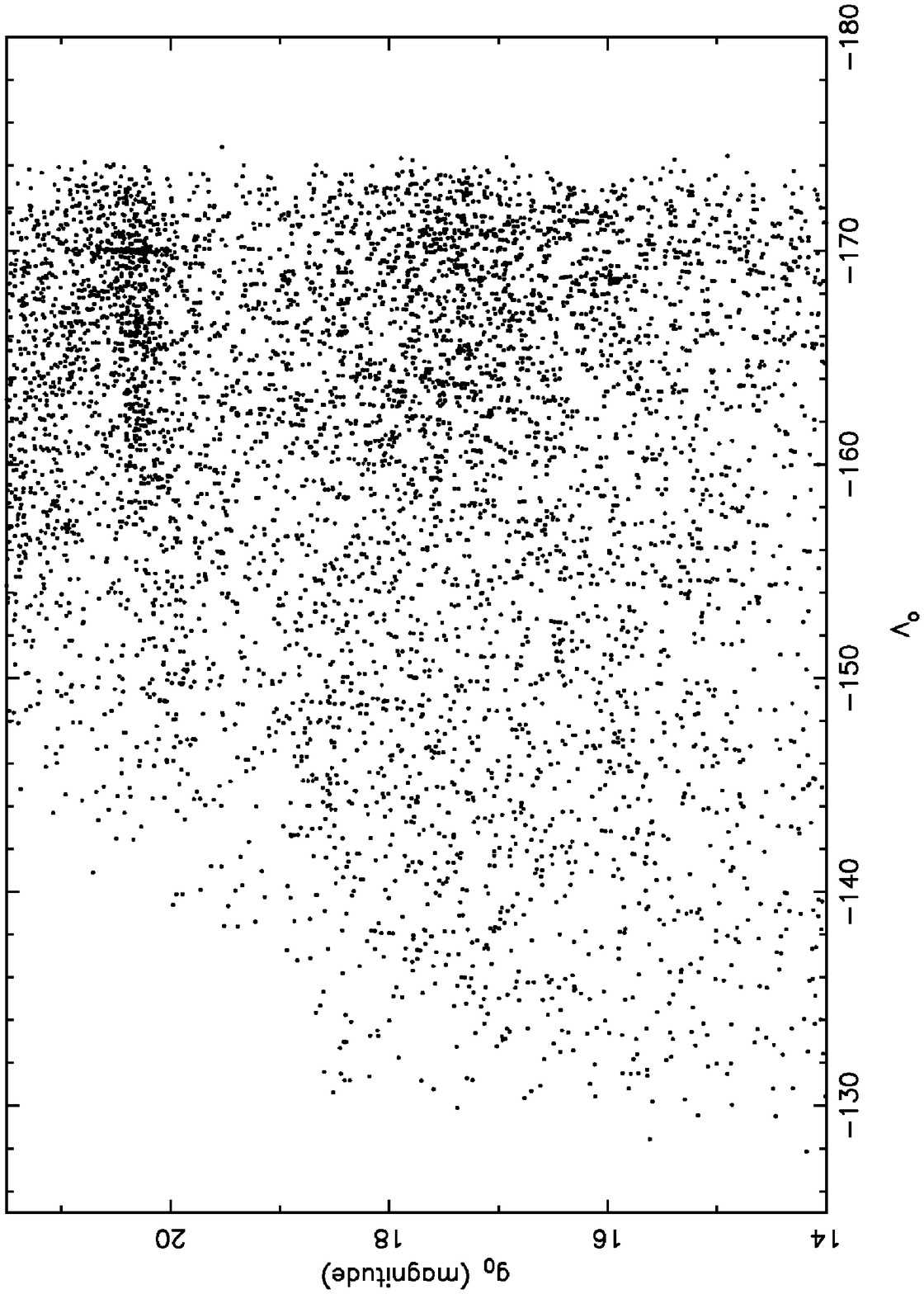}

\plotone{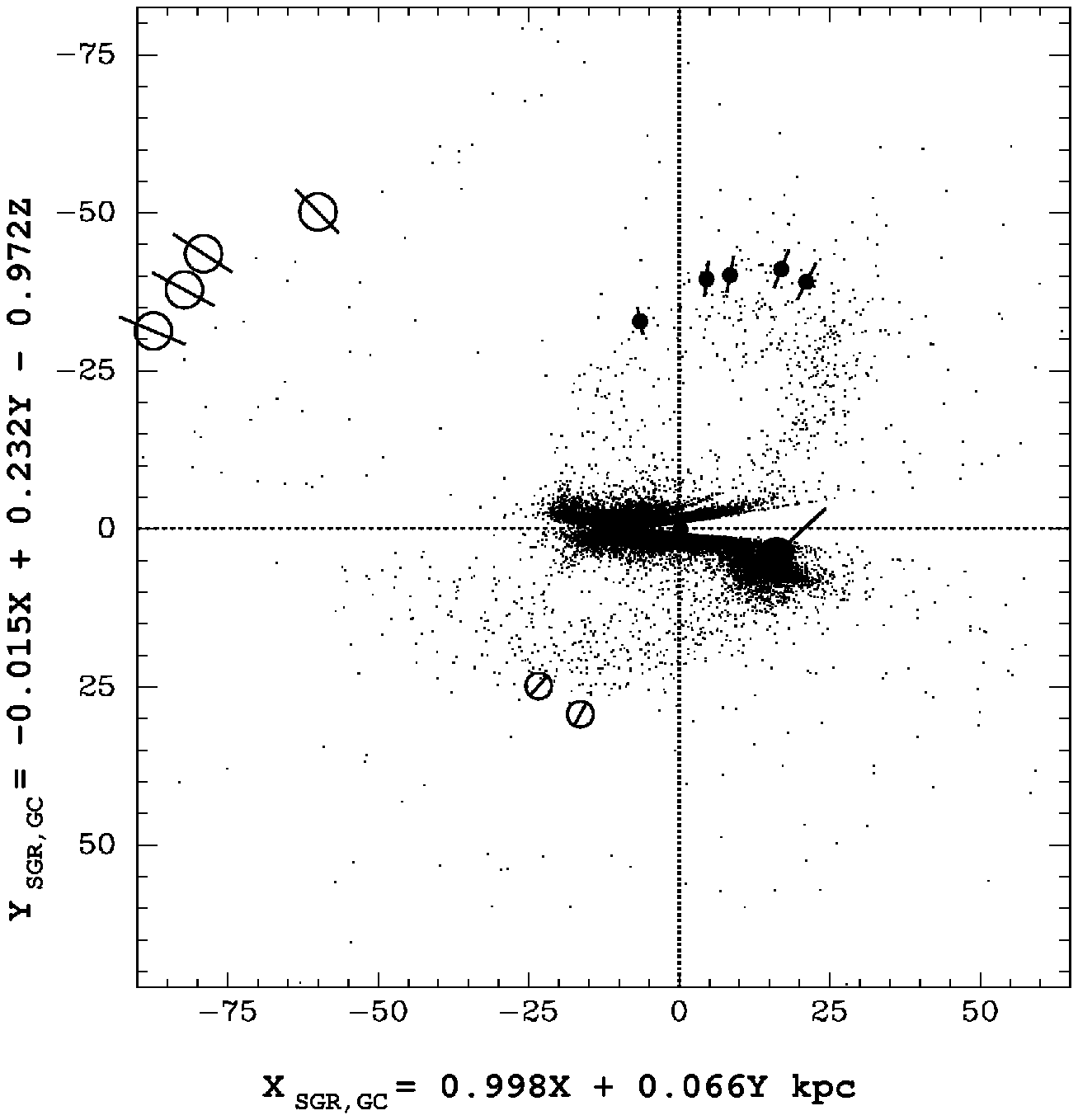}

\plotone{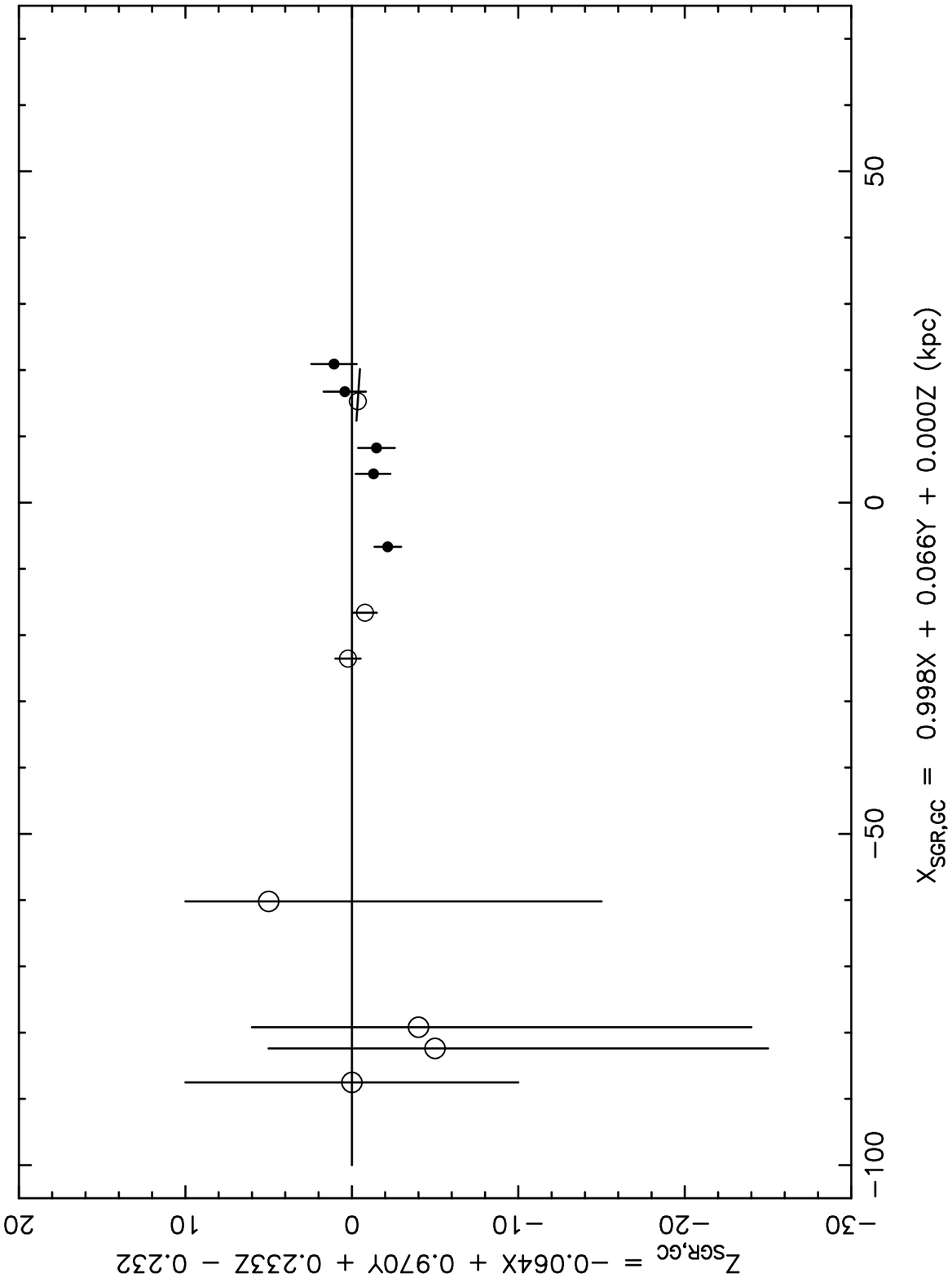}

\end{document}